\documentclass[%
aip,
amsmath,amssymb,%
author-year,%
twocolumn%
]{revtex4}

\usepackage{graphicx}
\usepackage{dcolumn}
\usepackage{bm}

\usepackage[utf8]{inputenc}
\usepackage[T1]{fontenc}
\usepackage{newtxtext}
\usepackage[varvw]{newtxmath}

\usepackage{color}
\usepackage{bm}
\usepackage{amsmath}
\usepackage{amsfonts}
\usepackage{amssymb}
\usepackage{bm}
\usepackage{dsfont}
\usepackage{float}

\usepackage{titlesec}
\newcommand{\intv}[1]{\ensuremath{\int #1 \mathrm d ^3 v}}

\renewcommand{\vec}[1]{\ensuremath{\bm{#1}}}
\newcommand{\mat}[1]{\ensuremath{\bm{\mathrm{#1}}}}

\newcommand{\parac}[1]{\ensuremath{#1_{z}}}
\newcommand{\perpc}[1]{\ensuremath{#1_{\perp}}}

\newcommand{\dn}[1]{\ensuremath{\mathrm d^{#1}}}
\newcommand*{\noaddvspace}{\renewcommand*{\addvspace}[1]{}}
\addtocontents{lof}{\protect\noaddvspace}
\newcommand{\ub}{\ensuremath{\hat{\vec z}}}

\begin{document}
\preprint{AIP/123-QED}

\title{Simulation of ion temperature gradient driven modes with 6D kinetic Vlasov code} 

\author{M. Raeth}
\email{Mario.Raeth@ipp.mpg.de}
\affiliation{Max Planck Institute for Plasma Physics, Boltzmannstr. 2, 85748 Garching, Germany}

\author{K. Hallatschek}
\affiliation{Max Planck Institute for Plasma Physics, Boltzmannstr. 2, 85748 Garching, Germany}

\author{K. Kormann}
\affiliation{Ruhr-Uni­ver­si­tät Bo­chum, Uni­ver­si­täts­stra­ße 150, 44780 Bo­chum, Germany}

\date{\today}

\begin{abstract} With the increase in computational capabilities over the last years it becomes possible to
	simulate more and more complex and accurate physical models. Gyrokinetic theory has been introduced in the 1960s
	and 1970s in the need of describing a plasma with more accurate models than fluid equations, but eliminating
	the complexity of the fast gyration about the magnetic field lines. Although results from current gyrokinetic
	computer simulations are in fair agreement with experimental results in core physics, crucial assumptions made in the
	derivation make it unreliable in regimes of higher fluctuations and stronger gradient, such as the tokamak
	edge. With our novel optimized and scalable semi-Lagrangian solver we are able to simulate ion-temperature
	gradient modes with the 6D kinetic model including the turbulent saturation. After thoroughly testing our
	simulation code against analytical computations and gyrokinetic simulations (with the gyrokinetic code GYRO),
	it has been possible to show first plasma properties that go beyond standard gyrokinetic simulations. This
	includes the explicit description of the complete perpendicular energy fluxes and the excitation of high
	frequency waves (around the Larmor frequency) in the nonlinear saturation phase. \end{abstract}

	\maketitle
    
\section{Introduction}
Kinetic models are capable of describing physical phenomena in a tokamak plasma on all scales from the size of the device ($\sim$1m) down to microscales of the order of the electron Larmor radius ($\sim 10^{-4}$m). However, until recent years, the computational capabilities did not allow the computation of the full 6D-kinetic equation for a magnetized plasma with the time resolution required to resolve the fast orbital motion of the articles around the magnetic field. To circumvent this problem, many gyrokinetic models have been developed \cite{littlejohn_hamiltonian_1981}, which reduce the dimensionality and the required time resolution by eliminating frequencies of the order of the Larmor frequency $\omega_{c}$.\\
Gyrokinetic simulations codes such as GENE \cite{jenko_massively_2000}, GS2 \cite{kotschenreuther_comparison_1995}, GYRO \cite{candy_high-accuracy_2016} or ORB5 \cite{jolliet_global_2007} are in good agreement with experiments in the core of fusion devices where only small perturbation amplitudes and gradients (in density and temperature) are present. Nonetheless, in regimes of large gradients and high fluctuation amplitudes, such as the plasma edge of a tokamak, the gyrokinetic approximations are debatable and at least those models based on a $\delta f$ approximation break down completely. Moreover, the limited frequency range in gyrokinetic simulations precludes the accurate representation of physical processes associated with high-frequency modes. To illustrate this point, Craddock et al. demonstrated the suppression of turbulence by ion Bernstein waves in their study \cite{craddock_theory_1994}.\\
With the rise of modern high-performance computing capabilities, it becomes more and more viable to simulate the full 6D kinetic equation. For this purpose, the massively parallel semi-Lagrangian code BSL6D for the Vlasov equation in 6D phase space has been developed \cite{kormann_massively_2019}. The code, BSL6D, is capable of simulating plasma turbulence across a broad range of frequencies, including those beyond the Larmor frequency, and with arbitrarily large fluctuation amplitudes.\\
Past efforts to venture into regimes beyond gyrokinetic theory include the drift-cyclotron model by Waltz and Deng \cite{waltz_nonlinear_2013} and a non-gyrokinetic magnetized plasma turbulence code developed by Sturdevant et al. \cite{sturdevant_implicit_2016,sturdevant_low_2017,miecnikowski_nonlinear_2018}. However, they either still use a reduced kinetic model, or are limited to a single toroidal model, effectively resulting in a 5D simulation. This work shows unprecedented 6D nonlinear kinetic simulations of the slab ITG instability, examining the energy fluxes inherent in the model.\\ 
In Section \ref{sec_simulation}, we provide a concise introduction to the physical model implemented in the code for slab Ion Temperature Gradient (ITG) simulations. Subsequently, we conduct a verification study in the linear regime, encompassing a comparison with the analytically derived dispersion relation (section \ref{sec_dispersion_relation}) and the computation of quasi-linear energy fluxes (section \ref{sec_energy_flux}). Finally, in section \ref{sec_non_linear_runs}, we present results from simulations capturing the nonlinear saturation phase of the ITG instability, comparing them to results obtained from the gyrokinetic code CGYRO. Intriguingly, the 6D kinetic simulations reveal that the saturation phase leads to the excitation of high-frequency ion Bernstein waves. We conducted our nonlinear studies employing both a Boussinesq approximation for the gradients (corresponding to the gyrokinetic description of the gradients) and a fully nonlinear treatment. This allows us to investigate the saturation phase under different conditions and gain a deeper understanding of the underlying physics.
The discovery of high-frequency ion Bernstein waves during the saturation phase challenges conventional understanding and opens new avenues for research into the nonlinear dynamics of plasma turbulence.

\section{6D semi-Lagrangian kinetic turbulence code for magnetized plasmas} \label{sec_simulation}
The Vlasov equation describes the motion of a plasma in presence of electromagnetic fields.
We consider ions in a constant and homogeneous magnetic field $\vec B = \hat{\vec z}$ represented by a 6D distribution function $f$, with an electric field $\vec E$ originating from the interaction with adiabatic electrons. The target of our 6D simulations is the ion kinetic equation in dimensionless variables ($\rho_i\; (\text{Ion Larmor radius}) = v_{\text{th}}\;(\text{Ion thermal velocity})=n\;(\text{Ion background density})=T \;(\text{Ion background temperature})=1$) 
\begin{equation}
  \partial_t f+\vec v\cdot\nabla f+(-\nabla\phi+\vec v\times\vec  \hat{\vec z})
  \cdot\nabla_{\vec v} f=0.\label{eq_vlasov}
\end{equation}
For the simulation of ion of temperature gradient (ITG) driven modes, the gradient is introduced in Boussinesq approximation. For this purpose, we assume that a background distribution function $f_0$ exists, which fulfills 
\begin{align}
	\vec v \cdot \nabla f_0 + \vec v \times \hat{\vec z} \cdot \nabla_v f_0 = 0.
\end{align}
This condition is met, when the background distribution is parameterized by an arbitrary function $g$ with $f_0(\vec r, \vec v) = f_0(\vec R + \vec \rho, \vec v) = g(\vec R, \perpc v, \parac v)$ (with $\parac v = \vec v \cdot \hat{\vec z}$, $\perpc v = |\vec v \times \ub|$ and $\vec R = \vec r - \vec \rho = \vec r + \vec v \times \vec \hat{\vec z} $ is the
location of the gyrocenter of a given particle, where $\vec r$ is the configurations space coordinate and $\vec\rho=\ub \times \vec v$ is the Larmor radius vector). The background distribution is chosen to resemble a Maxwellian with varying
temperature $T(\vec R)$, and thus reads
\begin{align}
	g_0(\vec R, \perpc v, \parac v) :=\left( \frac{1}{2\pi T(\vec R)}\right)^{\frac 32} \exp\left(-\frac{ [\perpc v ^2 +
		\parac v^2]}{2 T(\vec R)}\right).
\end{align}
When splitting the distribution function in the Vlasov equation into a perturbation $\delta f = f- g_0$ and the background $g_0$, one obtains 
\begin{align}
	\partial_t \delta f + \vec v\cdot \nabla \delta f + \left[-\nabla \phi+(\vec v\times \hat{\vec z})\right] \cdot \nabla_v \delta f =  \nabla \phi\cdot \nabla_v g_0,
\end{align}
where the gradient on the right-hand side is given by
\begin{align}
	\nabla_v g_0 &= \ub \times \nabla g_0 + \nabla_v g_0  =: {\vec v^*}g_0 + \nabla_v g_0\notag \\
	&= -g_0 \ub \times \left( \frac{ \nabla T}{T} \frac{3 -  (\perpc v ^2 + \parac v^2)}{2 }\right) +  \nabla_v g_0 .\label{eq_source_term}
\end{align}
The limit that the gradient length is infinite $T(\vec R) \rightarrow T = 1$ is taken, such that the background distribution is independent of $\vec R$. Thus,  $g_0$ is replaced with the homogeneous Maxwellian distribution $f_M = \frac{1}{(2\pi)^{\frac 32}}e^{-\frac {v^2}{2}}$. \\
For simplicity, we assume that the electrons are adiabatic $f_e = e^{-\phi}f_M \approx f_M - \phi f_M$ and the plasma is quasineutral $n_e=n$  (with the electron density $n_e$), the complete system of equation is given by  
\begin{align}
	\partial_t f + \vec v \cdot \nabla f + \left( - \nabla \phi + \vec v \times \ub \right)\cdot \nabla_v f = 
	\vec v^* \cdot \nabla \phi f_M, \label{eq_temperature_gradient_system} \\
	\phi = \delta n, \quad \quad \quad\vspace{2cm} \delta n = \int h \dn3v. \label{eq_field_equation}
\end{align}
For all presented results, the electron temperature has been chosen to be the same as the ion temperature $T_e = 1$.
The system is integrated in time using a semi-Lagrangian method \cite{sonnendrucker_semi-lagrangian_1999}, which combines the advantages of grid-based (Eulerian) and particle-based (Lagrangian) methods. In this method, the distribution function is represented as a discrete function on a grid. However, instead of explicitly computing partial derivatives (as in Eulerian methods), the distribution function is advected along the characteristics of the Vlasov equation. This is achieved by solving the characteristic equations for particle trajectories and interpolating the distribution function at their new positions. This approach circumvents the strict CFL (Courant-Friedrichs-Lewy) condition imposed on Eulerian methods, allowing for the selection of a larger time step. Additionally, the noise levels are significantly lower compared to particle methods, making the semi-Lagrangian method a more robust and efficient choice for simulating the dynamics of plasmas. Implementation details can be found in the work of Kormann et al. \cite{kormann_massively_2019}.

\section{Dispersion relation}
\subsection{Analytical description}
The response of the non-adiabatic distribution, given by $h = f - f_M + \phi f_M$, to a present electrostatic potential $\phi$ in the linearized system is given by the Gordeev integral \cite{gordeev_notitle_1952}. This computation yields an expression for the velocity distribution ($\vec{v} = (v_x, v_y, v_z)^T$) of a specific Fourier mode of the distribution function $h_k$, influenced by an electrostatic wave $\phi = \phi_k e^{i(\vec{k}\cdot\vec{r} - \omega t)}$ with frequency $\omega$ and wave vector $\vec{k}$ (where $\parac k = \vec{k}\cdot\ub$, $\perpc{\vec{k}} = \vec{k} - k_z \ub$, and $\alpha$ is the azimuth angle in velocity space with $\hat{\vec z} \times \vec k$ corresponding to $\alpha = \frac \pi 2$)  \cite{brambilla_kinetic_1998}
\begin{align}
  h_k =   \phi_k(\omega - \vec{k}\cdot\vec{v^*}) f_M\sum_{m,p\in\mathbb{Z}} \frac{J_m(\perpc k \perpc v)J_p(\perpc k \perpc v)e^{i(p-m)\alpha}}{\omega - k_z v_z - p}, \label{eq_anatz_distribution}
\end{align}
where $J_p(x)$ represents the Bessel function of the first kind, and $\vec v^*$ is defined in equation (\ref{eq_source_term}). The density response $n_k = \int h_k \mathrm d^3 v - \phi_k$ is given by
\begin{align}
  n_k =  &\int_{-\infty}^{\infty}\int_{0}^{\infty} \int_0^{2\pi}i (\omega -\vec k \cdot \vec v^*) \phi_k f_M\notag \\&\sum_{m,p \in
    \mathds Z} \frac{J_m(\perpc k \perpc v)J_p(\perpc k \perpc v) e^{i (p-m)\alpha}}{k_z v_z - \omega -p} \perpc v \mathrm
  d\alpha \mathrm d \perpc v \mathrm d \parac v - \phi_k. \label{eq_density_response_int}
\end{align}
To simplify the integration, the source term $\vec k \cdot \vec v^* f_M$, defined in equation (\ref{eq_source_term}) can be written in terms of a derivative with respect to the helping variable $\xi$ (assuming $\frac{\nabla T}{T}\propto \hat{\vec y}$)
\begin{align}
	\vec k \cdot \vec v^* f_M(\perpc v, \parac v) = k_y \frac{\nabla T}{T}\partial_\xi  f_M(\sqrt \xi \perpc v, \sqrt \xi \parac v), 
\end{align}
which is set $\xi=1$ after the derivative is calculated. The parallel velocity integral is resolved using the plasma dispersion function 
\begin{align}
	Z(x) = \frac{1}{\sqrt\pi} \int_{-\infty}^\infty \frac{e^{-t^2}}{(t-x)^2}\mathrm d t.
\end{align}
 After solving the perpendicular integral, we introduce $\Gamma_n(x^2) = e^{-x^2}I_n(x^2)$, resulting in the expression \cite{brambilla_kinetic_1998}
\begin{align}
    \frac{n_k}{\phi_k} = \left[\omega- k_y\frac{\nabla T}{T}\partial_\xi\right] \frac{\sqrt \xi}{|k_z|\sqrt{{2}}}\sum_{p \in \mathds Z}  Z\left(\frac{\omega+p}{|k_z|} \sqrt{\frac{\xi}{2}}\right)\Gamma_n\left(\frac{\perpc k^2}{\xi}\right)-1.\label{eq_density_response}
\end{align}
The density response is introduced in the quasi-neutrality equation which gives the electrostatic potential form equation (\ref{eq_field_equation}).
The resulting dispersion relation, given by solution $\omega(\vec k)$ to the equation
\begin{align}
0=\left[\omega - k_y\frac{\nabla T}{T}\partial_\xi\right]  \frac{\sqrt \xi}{|k_z|\sqrt{{2}}}\sum_{p \in \mathds Z}  Z\left(\frac{\omega+p}{|k_z|} \sqrt{\frac{\xi}{2}}\right)\Gamma_n\left(\frac{\perpc k^2}{\xi}\right)-2, \label{eq_dispersion_relation}
\end{align}
describes all modes for the 6D kinetic system. In the gyrokinetic limit $\omega\ll1$ and $k_z\ll 1$ all terms of the sum vanish except the $p=0$ contribution, simplifying the dispersion relation to 
\begin{align}
  0=\left[\omega - k_y\frac{\nabla T}{T}\partial_\xi\right] \frac{\sqrt \xi}{|k_z|\sqrt{{2}}}Z\left(\frac{\omega}{|k_z|} \sqrt{\frac{\xi}{2}}\right)\Gamma_0\left(\frac{\perpc k^2}{\xi}\right)-2. \label{eq_gyro_dispersion_relation}
\end{align}

\subsection{Verification of dispersion relation \label{sec_dispersion_relation}}
For the numerical verification of the code against the dispersion relation, a linear simulation has been conducted. The removal of the non-linear term from the Vlasov equation (\ref{eq_vlasov}) results in indefinite linear growth, simplifying the determination of growth rates.
In space, a box with dimensions $N_x = 128 \times 128 \times 8$ has been selected. The box size ensures the smallest perpendicular wave number is $k_{x,0}=k_{y,0} = 0.3$ ($L_x = L_y = \frac{20}{3}\pi$). The parallel length is determined by the wave number of the fastest-growing mode in the system, dependent on the temperature gradient. A temperature gradient of $\frac{\nabla T}{T} = 0.05$ has been employed, leading to an ideal parallel wave number of $k_{z} = \frac{1}{240}$ and, consequently, $L_z = 480\pi$. The velocity space is symmetric in all directions, with $v_{\mathrm{max}} = 4$ and $N_v = 32 \times 32 \times 32$. For interpolation, a 7th-order Lagrange interpolation has been applied to the velocity directions, and an 8th order for the spatial directions.

The simulation commences with a small white noise density perturbation, and growth rates are determined by fitting the time evolution of various Fourier modes of the density.
\begin{figure}
  \centering
  \includegraphics[width=0.4\textwidth]{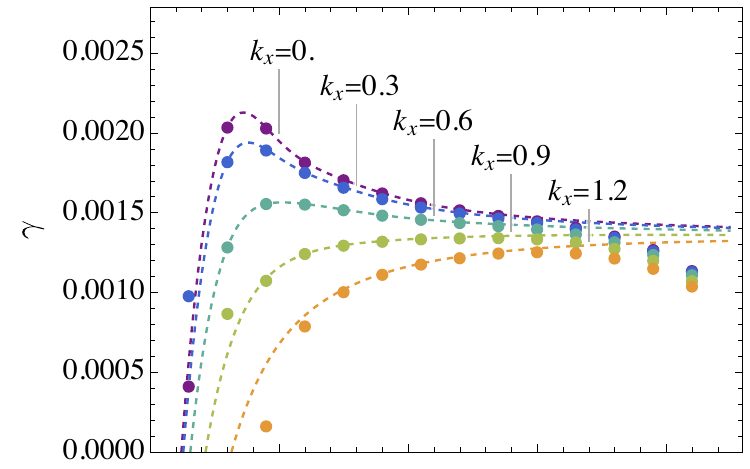}
  \includegraphics[width=0.4\textwidth]{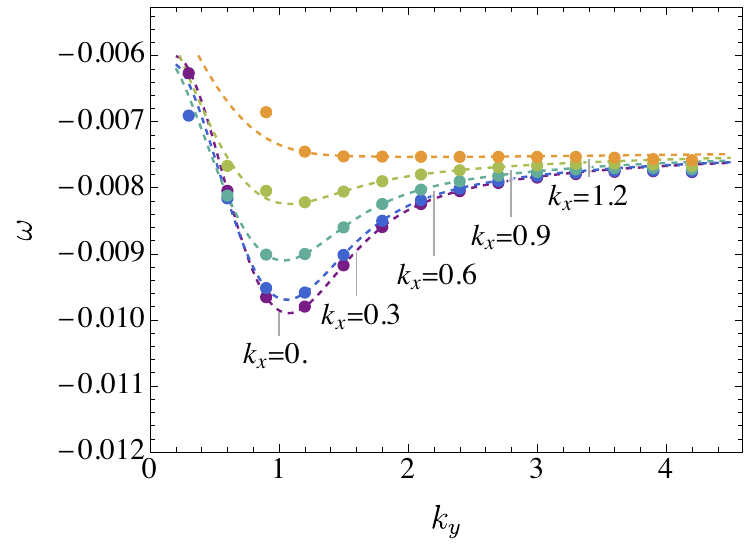}
	\caption{Comparison of the frequency (left) and the growth rate (right) for various modes with $k_z =
			\frac{1}{240}$ determined from linear simulations ($\bullet$) and computed by numerical root finding from the
		analytical dispersion relation (\ref{eq_dispersion_relation}) in solid lines \label{fig_dispersion_comparison}}
\end{figure}
Figure \ref{fig_dispersion_comparison} presents a comparison between the numerical results and the analytical dispersion relation. The code demonstrates excellent agreement for both the growth rate and frequency.
However, discrepancies are more pronounced for very small and larger wave numbers. For small $k_y$, the growth rates are minimal, or even negative, making the numerical computation challenging. As for larger perpendicular wave numbers, the numerical damping, attributed to the inherent diffusion of the Lagrange interpolation, is comparable to the ITG growth rates. Consequently, the code tends to underestimate the growth rate. This discrepancy can be mitigated by either enhancing the resolution or adopting a higher interpolation order \cite{rath_beyond_2023}.
\section{Contributions to the energy flux in the 6D system}
\subsection{Analytical description \label{sec_energy_flux}}
The total energy density $E$ of the system results from the sum of the kinetic energy of ions and the energy stored in the electrons
\begin{align}
  E &= \int \frac {v^2}2 f  \mathrm{d}^3v + \frac{n^2}2.
\end{align}
The time derivative of the total energy is computed using the Vlasov equation (\ref{eq_vlasov})
\begin{align}
  \partial_t E =& \int\frac {v^2}2 \partial_t f  \mathrm{d}^3v + n \partial_t n \notag \\
  =& - \int \frac {v^2}2 \vec v \cdot \nabla f  \mathrm{d}^3v\notag \\
  & -  \int\frac {v^2}2 (-\nabla \phi +\vec v \times \ub)\cdot \nabla_{\vec v} f  \mathrm{d}^3v\notag \\ &+n \partial_t n.
\end{align}
Applying integration by parts to the second integral yields
\begin{align}
  \partial_t E &= - \int \frac {v^2}2 \vec v \cdot \nabla f  \mathrm{d}^3v -  \int \vec v\cdot ( \nabla \phi) f  \mathrm{d}^3v+n \partial_t n
\end{align}
which farther can be modified to be expressed as a divergence of the entire integral
\begin{align}
  \partial_t E &= -\nabla \cdot \left(\int \frac {v^2}2 \vec v  f \mathrm{d}^3v\right) - \nabla \cdot  \left(\phi\int \vec v   f  \mathrm{d}^3v\right)\notag \\ &+ \phi \int \vec v\cdot  \nabla f  \mathrm{d}^3v+n \partial_t n.
\end{align}
By introducing $\partial_t n = -\nabla \cdot \Gamma :=-\nabla \cdot \int \vec v  f  \mathrm{d}^3v$ and utilizing the field equation (\ref{eq_field_equation}), the expression can be succinctly summarized as
\begin{align}
	\partial_t E &=  -\nabla \cdot \vec Q - \nabla \cdot (\phi \vec \Gamma). \label{eq_2nd_moment_vlasov}
\end{align}
Apart from $ \vec Q =\frac 12 \int \vec v v^2 f \dn3v$, an additional energy flux $\vec S = \phi \vec \Gamma$ appears which, can be identified as a Poynting like energy flux. The energy flux in the gyrokinetic system is completely described by the $\vec E \times \vec B$ heat flux. Thus, in the gyrokinetic limit, the energy flux in the 6D system has to be equal the $\vec E \times \vec B$ heat flux.
The various contributions to the energy flux $\vec Q = \int \frac {v^2}2 f \dn 3 v$ are obtained by computing the time derivative with the Vlasov equation (\ref{eq_vlasov}).
The Lorentz force term of the Vlasov equation is altered by partial integration
\begin{align}
  \int& \frac{v^2}{2} \vec v\left[ (-\nabla \phi + \vec v \times \ub) \cdot \nabla_{\vec v} f \right] \dn 3 v \notag \\=& \int \left(\nabla \phi \cdot \vec v\right)  \vec v f \dn 3v + \int \frac{v^2}2 \nabla \phi f \dn 3 v - \int (\vec v\times \ub) \frac{v^2}{2} f\dn 3 v
\end{align} 
After introducing the expression for kinetic energy density $\epsilon =\int \frac{v^2}2 f \dn3v $, the stress tensor $\mat \Pi  = \int \vec v \vec v f \dn3v$,  $\mat \Pi^* = \int \frac{v^2}{2}\vec v \vec v f \dn3v$ and computing the cross product with the magnetic field $\partial_t \vec Q \times \hat{\vec z}$, the Grassmann identify can be applied to obtain an expression for the perpendicular energy flux
\begin{align}
  \partial_t{ \vec Q}\times \hat{\vec z}  = -&\perpc {\vec Q}- (\nabla \cdot \mat \Pi^*)\times \hat{\vec z}\notag \\&-  (\mat \Pi\cdot \nabla \phi) \times \hat{\vec z} - \nabla \phi\times \hat{\vec z} \epsilon.
\end{align}
For low frequency waves $\omega\ll 1$, the time derivative can be neglected $\partial_t \vec Q \approx 0$ and without boundary contributions, the flux surface ($y-z-$plane, $\langle \cdot \rangle := \frac{1}{L_y L_z}\int \cdot \mathrm d y \mathrm dz$) average results in
\begin{align}
  \langle\perpc{\vec Q}\rangle =  {-\langle\nabla \phi\times  \hat{\vec z}   \epsilon\rangle}   -\langle(\mat \Pi \cdot\nabla \phi) \times \hat{\vec z}\rangle.
\end{align}
The overall change in energy density can be written as 
\begin{align}
 \langle\partial _t E \rangle =\nabla \cdot \left(\langle \nabla \phi\times  \hat{\vec z}   \epsilon  \rangle+   \langle(\mat \Pi \cdot\nabla \phi) \times \hat{\vec z} \rangle- \langle\vec S\rangle \right).\label{eq_total_energy_flux}
\end{align}
In conclusion, the mean perpendicular energy flux consists of three contributions
\begin{enumerate}
\item Poynting flux $\vec S = \phi \vec \Gamma $
\item $\vec E \times \vec B$ - energy flux $\vec Q^{\vec E \times \vec B} =-   (\nabla \phi\times  \hat{\vec z}   )\epsilon $
\item Stress tensor induced energy flux $\vec Q^{\mat \Pi} =(\mat \Pi \cdot\nabla \phi) \times \hat{\vec z}  $
\end{enumerate}
This depiction of the energy flux offers two significant advantages. Firstly, it provides a direct mean of calculating the individual contributions to the energy flux from the code. In the gyrokinetic limit, the Poynting flux $\vec{S}$ and the stress-induced heat flux $\vec{Q}^{\mat \Pi}$ nullify each other, leaving the $\vec{E} \times \vec{B}$-heat flux as the sole remaining form of energy transport. The equivalency between these two energy fluxes serves as a useful test for verifying the accuracy of the energy flux in the gyrokinetic regime. Secondly, it facilitates the direct computation of contributions to the energy flux, rendering it less susceptible to gyro-oscillations and, consequently, easier to calculate.\\
In the following, we analytically compute the contributions to the energy flux for a given solution of the dispersion relation.\\

\paragraph{1. The Poynting flux} $\vec S = \phi \vec \Gamma$ is given by the particle flux multiplied by the electrostatic potential. The derivation of the particle flux, denoted by $\vec \Gamma = \int \vec v f \mathrm{d}^3v$, follows the same approach as for the density response in equation (\ref{eq_density_response}) using the ansatz for the distribution function from equation (\ref{eq_anatz_distribution}).\\
For the perpendicular flux, computations are carried out independently in the two directions $(v_1,v_2) = \perpc v (-\sin \alpha, \cos \alpha)$, where $\alpha$ is the azimuth angle in velocity space, ensuring that the angle of $\hat{\vec z} \times \vec k$ corresponds to $\alpha = \frac{\pi}{2}$. For a more general calculation, the integral is evaluated for an arbitrary complex Fourier mode $P_l = \int \perpc v e^{ila} f \mathrm{d}^3v$. We recall the definition of the non-adiabatic perturbation of the distribution function $h = f - f_M + \phi f_M$, which allows us to compute the integral $P_l$ from equation (\ref{eq_anatz_distribution})
\begin{align}
	P_l&= \int \perpc vh_k e^{il\alpha} \mathrm d^3 v \notag\\ &= \int  \left[\omega -\vec k \cdot \vec v^*\right] \phi_k f_M
	\notag \\
  &\hspace{0.25cm}\times\sum_{m,p \in \mathds Z} \frac{J_m(\perpc k \perpc v)J_p(\perpc k \perpc v) e^{i (p-m)\alpha}}{\omega- k_z v_z  -p}
	\perpc v e^{il\alpha}\dn 3 v, 
\end{align}
and subsequently assemble the particle flux $\vec \Gamma$.
The double sum simplifies when the angle integral is evaluated with $\int_0^{2\pi} exp({i (p-m+l)\alpha})\mathrm d \alpha = 2\pi \delta_{m,p+l}$ which leads to the expression
\begin{align}
	P_l=2\pi \phi_k  &\int_{-\infty}^{\infty}\int_{0}^{\infty} \left[\omega -\vec k \cdot \vec v^*\right] f_M \notag\\&\sum_{p\in \mathds Z}
	\left[ \frac{J_{p+l}(\perpc v \perpc k) J_p(\perpc v \perpc k)}{(\omega - k_z v_z -p )}\right]\perpc v^2 \mathrm d
	\perpc v \mathrm d \parac v. \label{eq_harmonic_response}
\end{align}
The velocity integrals are solved numerically for a given set of parameters and the respective solution ($\omega,\vec k$) of the dispersion relation (\ref{eq_dispersion_relation}). When computing the flux in a spatial direction, the
definition of the velocity angle $\alpha$ has to be kept in mind. The angle is defined such that $\vec k \cdot \hat{\vec z}$
corresponds to $\alpha = \frac \pi 2$. Thus, the fluxes
\begin{align}
	\Gamma_1  &=\frac i 2 \left( P_{1} - P_{-1}\right),\\
	\Gamma_2 & =\frac 1 2 \left( P_{-1} + P_{1}\right),
\end{align}
point in the direction perpendicular (for $\Gamma_1$) and parallel (for $\Gamma_2$) of the wave vector. In the case of $\vec k = k \hat{\vec y}$, $\Gamma_1$ and $\Gamma_2$ corresponds to the $x$ and $y$ direction respectively. For a general wave vector, the fluxes need to be combined to compute the correct fluxes in $x$ and $y$ direction. The resulting velocity integrals
\begin{align}
	\frac{\Gamma_1}{\phi_k}=\Xi i  \pi   \int_{-\infty}^{\infty}&\int_{0}^{\infty} \mathrm d \perpc v \mathrm d \parac v\perpc v^2f_M(\sqrt \xi v_z, \sqrt \xi \perpc v)\notag\\&
	\times\left[\sum_{p\in \mathds Z}\left(\frac{J_{p+1}(\perpc v \perpc k) J_p(\perpc v \perpc k)}{ \omega - k_z v_z-p}\right
	.\right . \notag \\ &\hspace{1cm}\left.\left.-\frac{J_{p-1}(\perpc v \perpc k) J_p(\perpc v \perpc k)}{\omega - k_z
		v_z-p}\right)\right]  ,\notag \\
    \frac{\Gamma_2}{\phi_k}=\Xi  \pi    \int_{-\infty}^{\infty}&\int_{0}^{\infty} \mathrm d \perpc v \mathrm d \parac v\perpc v^2f_M(\sqrt \xi v_z, \sqrt \xi \perpc v)\notag\\&
  \times  \left[\sum_{p\in \mathds Z}\left(\frac{J_{p-1}(\perpc v \perpc k) J_p(\perpc v \perpc k)}{ \omega - k_z v_z-p}\right
    .\right . \notag \\ &\hspace{1cm}\left.\left.+\frac{J_{p+1}(\perpc v \perpc k) J_p(\perpc v \perpc k)}{\omega - k_z
      v_z-p}\right)\right], \label{eq_particle_flux_response}  
\end{align}
where $\Xi = \left[\omega- k_y \left(\frac{\nabla T}{T}\partial_\xi\right)\right]$, are then computed numerically.
\paragraph{2. The $\vec E \times \vec B$ heat flux} $\perpc{\vec Q}^{E\times B}  =  {-\nabla \phi\times  \hat{\vec z}   \epsilon}  $ can be computed analogously to the Poynting flux.  In Fourier space, the flux can be written as ${\perpc{ {\vec Q}}^{\vec E \times \vec B}  =  i   \phi \vec k\times \hat{\vec z} \epsilon_k}$, where the energy density response
$\epsilon_k = \int \frac{v^2}{2}h_k \dn3v$ is computed from the distribution function (\ref{eq_anatz_distribution}) resulting in 
\begin{align}
	&\epsilon_k = \phi_k\left[\omega \partial_\xi \Psi- \ub \times \frac{ \nabla T}{T} \partial_\xi\left(\xi^{\frac 32} 
	\partial_\xi \Psi\right)\right], \text{ with }\\ \notag
	&\hspace{1.5cm}\Psi =  \frac{1}{|k_z|}\sqrt{\frac \xi 2} e^{-\frac{\perpc k^2}{\xi}}\sum_{p \in \mathds Z} 
	Z\left(\frac{\omega+p}{|k_z|} \sqrt{\frac{\xi}{2}}\right)I_p\left(\frac{\perpc k^2}{\xi}\right)
	\label{eq_energy_response}.
\end{align}
\vspace{2cm}
\paragraph{3. The stress induced energy flux} is computed from the stress tensor response $	\mat \Pi_k = \int \vec v 
\vec v h_k \mathrm d^3 v$. When the velocity components $(v_1,v_2,v_z) = (-\perpc v \sin \alpha, \perpc v \cos \alpha, v_z)$ are introduced, the expression for the energy flux component reads
\begin{widetext}
\begin{align}
	\perpc{ {\vec Q}}^{\Pi} &=i\phi_k \left[(\mat \Pi_k \cdot\vec k) \times \hat{\vec z}\right]   =i   \phi  \intv{\frac 12\left(\begin{array}{c}
			{k_y} \perpc v ^2 - k_y \perpc v^2 \cos(2\alpha)+ 2 {k_z}  \perpc v\parac v \sin \alpha + k_x \perpc v^2
			\sin(2\alpha)  \\ -k_x \perpc v^2 - k_x \perpc v^2 \cos(2\alpha)- {k_y} \perpc v^2   \sin (2\alpha) - 2 {k_z}  \parac v
			\perpc v \sin \alpha \\0
		\end{array}  \right)f}. 
\end{align}
\end{widetext}
The energy flux is assembled using the response derived in equation (\ref{eq_harmonic_response}). The
velocity integrals have been computed numerically for the given parameters. 

\subsection{Verification of energy fluxes \label{sec_comparison_fluxes}}
In the numerical testing, the contribution to the energy flux derived in section \ref{sec_energy_flux} are computed from the numerically determined particle flux $\vec \Gamma = \int \vec v f \mathrm{d}^3v$, the energy density  $\epsilon = \int \frac{v^2}{2}f \mathrm{d}^3v$, and the components of the stress tensor $\mathbf{\Pi} = \int \vec v \vec v f \mathrm{d}^3v$. Subsequently, these numerical results are compared to the analytical descriptions for the individual contributions to the total energy flux as derived in equations (\ref{eq_total_energy_flux}). The analytical calculations involve the introduction of the particle flux (\ref{eq_particle_flux_response}), energy density (\ref{eq_energy_response}), and the stress tensor response, as described in equation (\ref{eq_total_energy_flux}).\\
\begin{figure}[H]
	\hspace{-0.5cm}
	\includegraphics[width=0.25\textwidth]{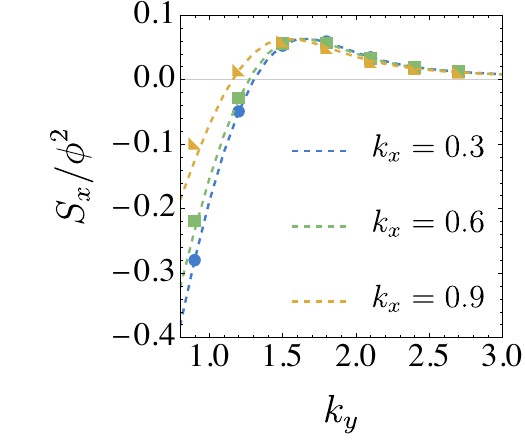}
  \includegraphics[width=0.25\textwidth]{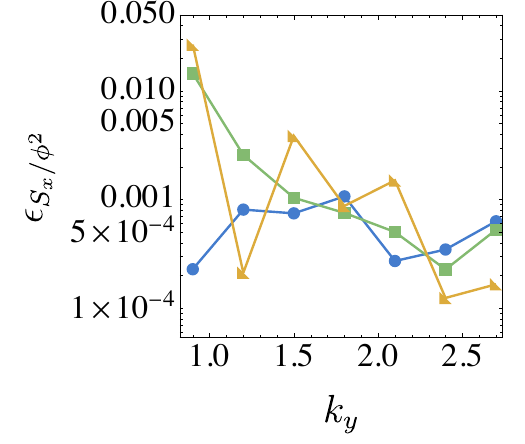}

  \hspace{-0.5cm}
  \includegraphics[width=0.25\textwidth]{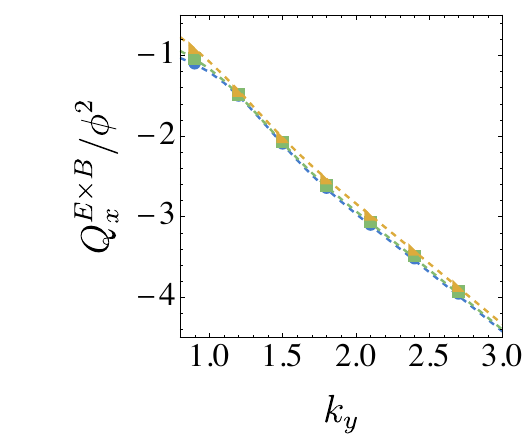} 
  \includegraphics[width=0.25\textwidth]{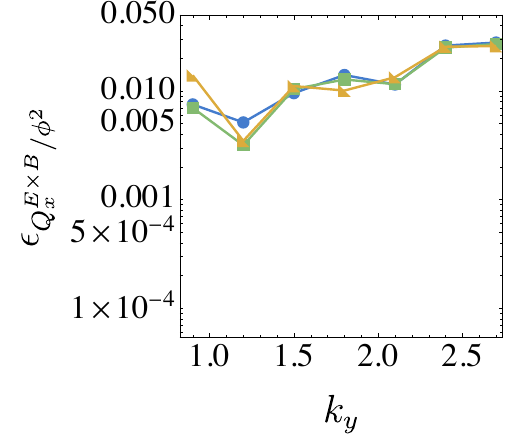}

	\hspace{-0.5cm}
  \includegraphics[width=0.25\textwidth]{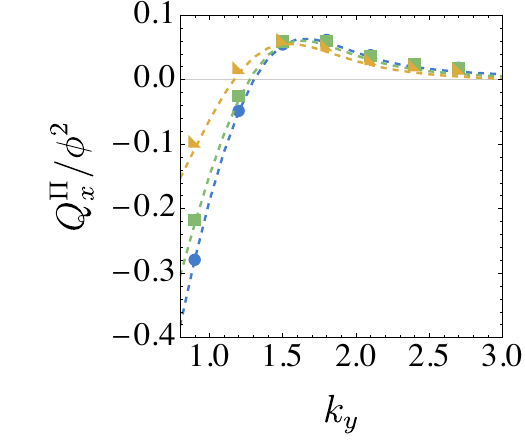}
  \includegraphics[width=0.25\textwidth]{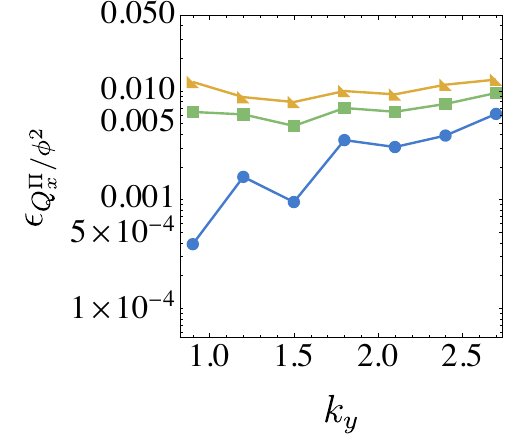}
  \caption{Comparison (left column) of energy fluxes (Poynting flux
  $\vec S$, $\vec E\times \vec B$ heat flux $\vec Q ^{\vec E\times \vec B}$, and stress induced heat flux $\vec Q^{\Pi}$) opposite to the temperature gradient for various wave numbers between (1) results from numerical
  simulations ($\bullet$) and (2) analytical computations (-). Respective deviation from the analytical results are shown in the second column \label{fig_flux_comparison} }
\end{figure}
In the simulations detailed in section \ref{sec_dispersion_relation}, where the growth rate and frequency of various modes have been determined, a linearized model is used. In the linearized limit, the velocity advection reduces to $\nabla \phi \cdot \nabla_v f \approx - \nabla \phi \cdot \vec v f_M$ to remove nonlinear effects. However, while this simplified model accurately reproduces the particle flux, higher moments of the distribution function are not faithfully represented. For these simulations, the full advection of the background distribution by the electric field must be considered. To prevent saturation, a very small initial perturbation amplitude (here,  $\delta n\sim 10^{-8}$) must be chosen. All other parameters remain consistent with the simulation of the dispersion relation in Section \ref{sec_dispersion_relation} ($N = 128 \times 128 \times 8 \times 32 \times 32 \times 32$, $L_x = L_y = \frac{20}{3}\pi$, $L_z = 480\pi$, $\Delta t = 0.03$), and the distribution function is initialized with a white noise density perturbation.\\   
The results for the contributions of the energy flux are shown in figure \ref{fig_flux_comparison} in comparison to the analytically obtained results together with the absolute error in the left column. 
The $\vec E \times \vec B$ heat flux is an order of magnitude larger than the other two fluxes. For large wave numbers, the $\vec E \times \vec B$ heat flux is proportional to the $y$-component of the wave vector $k_y$. The values of the Poynting flux and the stress induced heat flux are nearly identical, as anticipated, given that the gradient has been selected close to the gyrokinetic limit. In gyrokinetic theory, the entire energy flux is determined by the $\vec E \times \vec B$ heat flux.\\
The code results demonstrate strong agreement with analytical predictions. Absolute errors are less than 10\% for both the pointing flux and the stress-induced heat flux while the relative error of the $\vec{E} \times \vec{B}$ heat flux is approximately 1\%.\\

\section{Non-linear ITG simulations \label{sec_non_linear_runs}}
\subsection{Comparison with gyrokinetic simulation}
\begin{figure}[H]
    \includegraphics[width=0.32\textwidth]{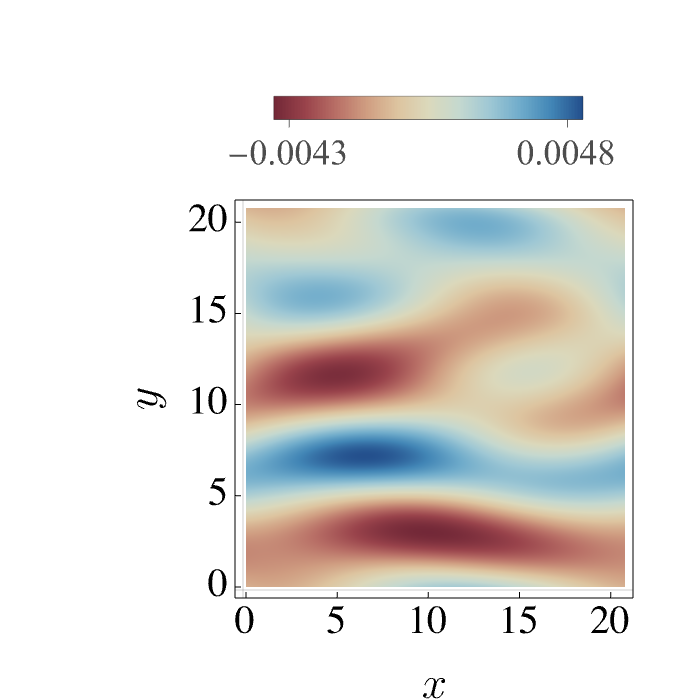}
    \includegraphics[width=0.32\textwidth]{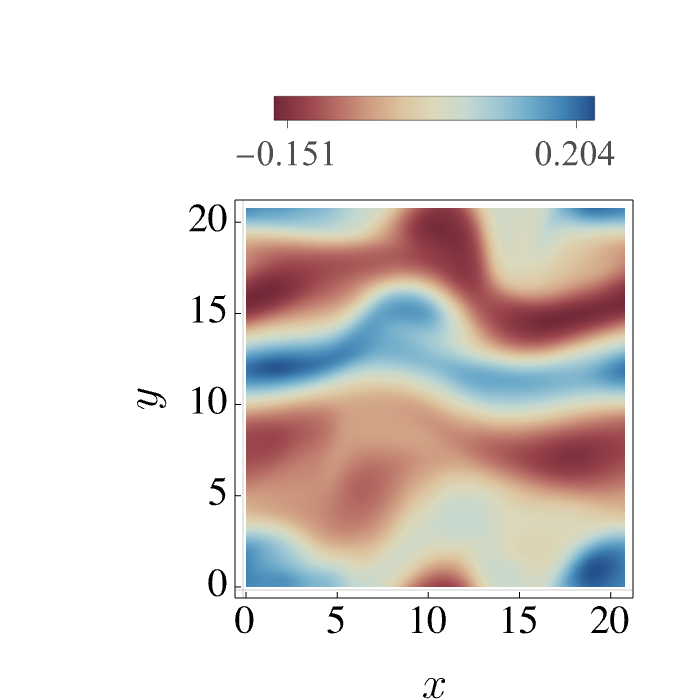}
    \includegraphics[width=0.32\textwidth]{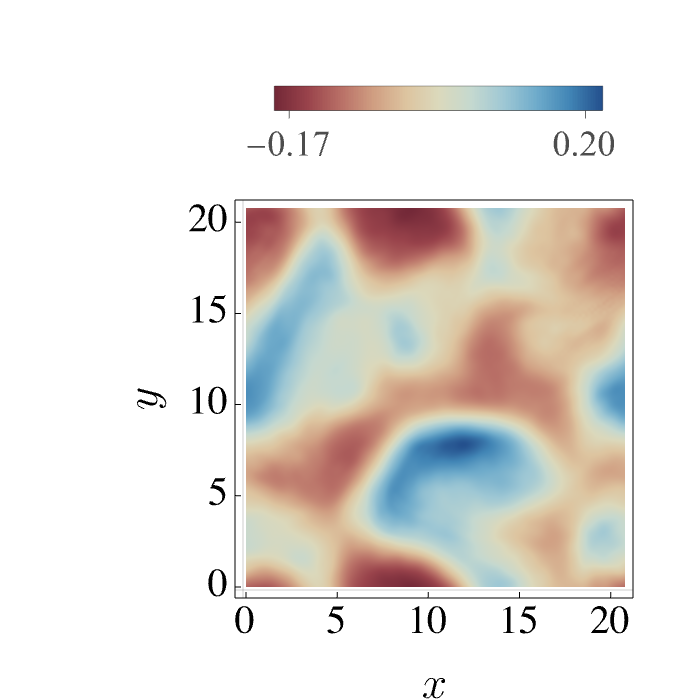}
    \caption{ Snapshots of the particle density in the $x$-$y$-plane (with fixed $z$) for various points in time
    \label{fig_snapshots_lin_itg} }
  \end{figure}

Having demonstrated the effective performance of the simulations in the linear regime, we have delved into the investigation of the nonlinear saturation of the ITG instability. For this purpose, a simulation similar to section \ref{sec_comparison_fluxes} has been conducted. However, the grid size has been adjusted from $N=128\times128\times 8$ to $N= 64\times 64\times 64$, specifically targeting the necessity for a higher resolution in $z$-direction due to the emergence of high-frequency waves during the nonlinear phase. This adjustment ensures that the various $x-y-$slices remained connected.\\
Figure \ref{fig_snapshots_lin_itg} displays snapshots of the density perturbation in the $x-y$ plane during three distinct phases of the simulation. In the first plot, a distinct mode structure is evident, characteristic of a slab ITG instability where the system is dominated by modes with $\vec{k} \perp \nabla T$ (in this case, $\nabla T/|\nabla T| = \hat{\vec{x}}$). The growth rates decrease rapidly as the wave number parallel to the temperature gradient increases. The second snapshot depicts the density during the transition from the linear to the nonlinear phase. The third image represents the perturbation in a fully saturated turbulent state.\\
The nonlinear saturation of the simulation is compared to the gyrokinetic code CGYRO \cite{candy_high-accuracy_2016}. To achieve this, a simulation with the same spatial parameters (length and resolution) is conducted. In gyrokinetic codes, the velocity space is not fully resolved as in 6D kinetic simulations but is parameterized by two coordinates: the kinetic energy, $\epsilon = [0, \epsilon_{\text{max}}]$, and the cosine of the pitch angle, $\xi \in [-1, 1]$. The pitch angle characterizes the ratio between parallel and perpendicular velocity. The chosen parameters for the simulation are $\epsilon_{\text{max}} = 8$, $N_\epsilon = 12$, and $N_\xi = 32$. This simulation is initiated with a random noise density perturbation having the same amplitude as the kinetic simulation.\\
\onecolumngrid

\begin{figure}
  \centering
  \includegraphics[width=0.4\textwidth]{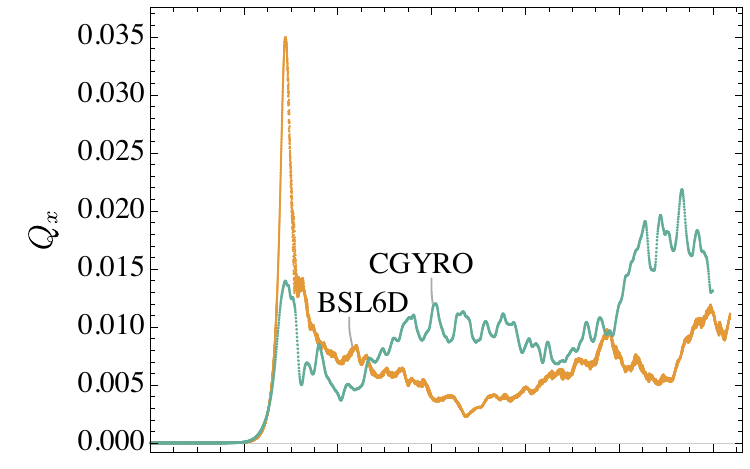}
  \includegraphics[width=0.4\textwidth]{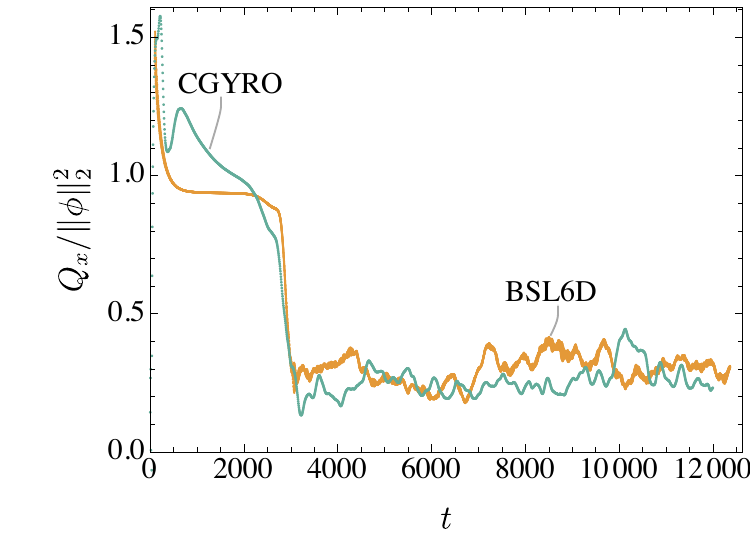}
  \caption{Comparison between BSL6D and gyrokinetic simulation (CGYRO) for the heat flux (a) and the ratio from heat flux
  and electrostatic potential \label{fig_itg_cgyro}}
\end{figure}

\twocolumngrid
For this comparison, emphasis will be placed on the heat flux induced during both the linear and nonlinear phases. Section \ref{sec_energy_flux} has established that the sole heat flux present in gyrokinetic theory is the $\vec E \times \vec B$ heat flux. Therefore, this serves as the metric for comparison. 
In Figure \ref{fig_itg_cgyro} (top), the heat flux parallel to the gradient is depicted for the two simulations. Following a linear phase (up to $t = 3000$), during which the flux experiences exponential growth, the energy flux reaches saturation. The saturation levels are similar for both simulations; however, in both cases, the levels are susceptible to significant fluctuations. This is partially attributed to the small domain size. The smallest wave number in the system is $k_0=0.3$, whereas the dominant growing mode has a wave number $k_y=0.9$. Consequently, only three wavelengths of the dominant mode fit into the system. Increasing the domain size would lead to a stronger averaging and thus, a more stable saturation level. In addition to expanding the domain, a longer runtime would enhance the comparison by providing more data and enabling a statistical analysis. However, due to the substantial computational cost associated with the 6D kinetic simulation, this option has been foregone, and the obtained results are considered satisfactory.

For further verification, we conducted a comparison of the ratio between the heat flux and the $L_2$-norm of the electrostatic potential, as depicted in Figure \ref{fig_itg_cgyro} (bottom). For the dominant mode, with $\vec k=(\frac 9{10},0,\frac{1}{240})$, a ratio of approximately 0.9 is anticipated, as indicated in Figure \ref{fig_flux_comparison} (2nd row, left), which is in close agreement with the simulated results.  Following the nonlinear phase, the ratio decreases to approximately 0.25, a consistent result for both models. In summary, a very good agreement is observed between the two simulations.

\begin{figure}
  \centering
  \includegraphics[width=0.4\textwidth]{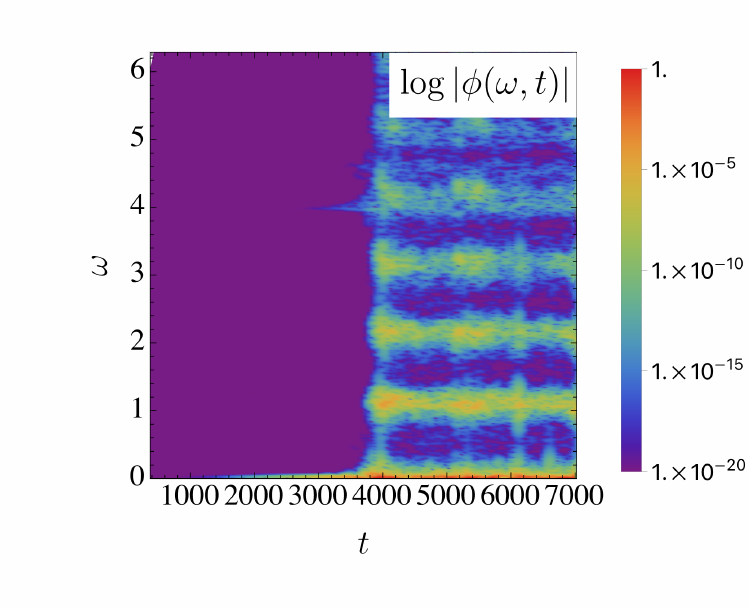}
  \caption{Spectrogram of electrostatic potential amplitude $\phi(\omega,t)$ \label{fig_spectrogram}}
\end{figure}

A notable distinction between the two simulations is evident. When examining the system in the turbulent state, one observes the presence of high-frequency oscillations with a frequency closely aligned with the Larmor frequency. The existence of such high-frequency oscillations in the saturation of nonlinear ion temperature gradient simulations has been previously acknowledged in the literature \cite{miecnikowski_nonlinear_2018}. However, a comprehensive description, let alone an investigation into their excitation, is still lacking. Figure \ref{fig_spectrogram} displays a spectrogram of the electrostatic potential. The ITG intensity is visible at the bottom of the figures with frequencies close to zero. Modes with frequencies close to the harmonics of the Larmor frequency are excited during the nonlinear saturation phase (at $t\sim 3500$). The modes are clustered in frequency bands close to the harmonics with a mean frequency slightly larger than the harmonic. The distribution of the high frequency waves suggest the presence of ion Bernstein waves (IBWs) \cite{candy_high-accuracy_2016}. More precisely neutralized IBWs are electrostatic waves in the ions with the presence of adiabatic electrons. Various mechanisms have been proposed to explain the instability of Ion Bernstein Waves (IBWs) \cite{raeth_high_2023,yoon_bernstein_2014,noreen_ion_2019}. Our analysis indicates that the excitation is triggered by local negative velocity space gradients of the distribution function, induced by the temperature gradient source term \cite{rath_beyond_2023}. These findings underscore the necessity of a nonlinear treatment of the temperature gradient in the full-f 6D model.  

\subsection{Nonlinear treatment of gradients}
This section transitions from a treatment of the gradients in a local limit, using the Boussinesq approximation to a nonlinear approach for handling gradients in our simulation code. The chosen initial condition in the simulation is designed to allow the distribution function to exhibit a density and temperature profile
\begin{align}
	f_0(\vec r, \vec v) = \frac{n(\vec R)}{(2\pi T(\vec R))^{\frac 32}}e^{-\frac{v^2}{2T(\vec R)}}.
\end{align}
The background profiles \(n(\vec r - \vec \rho)\) and \(T(\vec r - \vec \rho)\) are defined in gyrocenter coordinates \(\vec R := \vec r - \vec \rho\) (where \(\vec \rho := \ub \times \vec v\) is the Larmor radius vector) to establish a background that does not oscillate with the Larmor frequency. To simplify the treatment of boundary conditions, the profiles are periodically set up using a sine-profile in the \(x\)-direction
\begin{align}
	n(\vec R) &= 1 + \kappa_n \sin( k_0 \left(x -  v_y\right)),\\
	T(\vec R) &= 1 + \kappa_T \sin( k_0 \left(x -  v_y\right)) \label{eq_temperature_profile}.
\end{align}
To prevent the background density gradient from generating an electric field, all modes with wave numbers parallel to the gradient are removed from the electrostatic potential by subtracting the flux surface average over the \(y-z\)-plane
\begin{align}
	\phi = n - \langle n\rangle_{y,z} = n - \frac{1}{L_y L_z} \int n  \mathrm{d} y  \mathrm{d} z.
\end{align}
For the simulation, the parameters $\kappa_n = 0$, $\kappa_T = 0.5$, and $k_0 = 0.2$ have been chosen, resulting in a temperature gradient $\max_{x\in [0,L_x]}\frac{\partial_x T(x)}{T(x)} = 0.115$, where $L_x$ represents the box length in the $x$-direction. The simulation has been conducted on a box with a length of $L=10\pi \times \frac{5}{2} \pi \times 240\pi$, $N=128\times32\times16\times32\times32\times16$ and $\delta t = 0.02$).
\begin{figure}
	\centering
		\includegraphics[width=0.8 \linewidth]{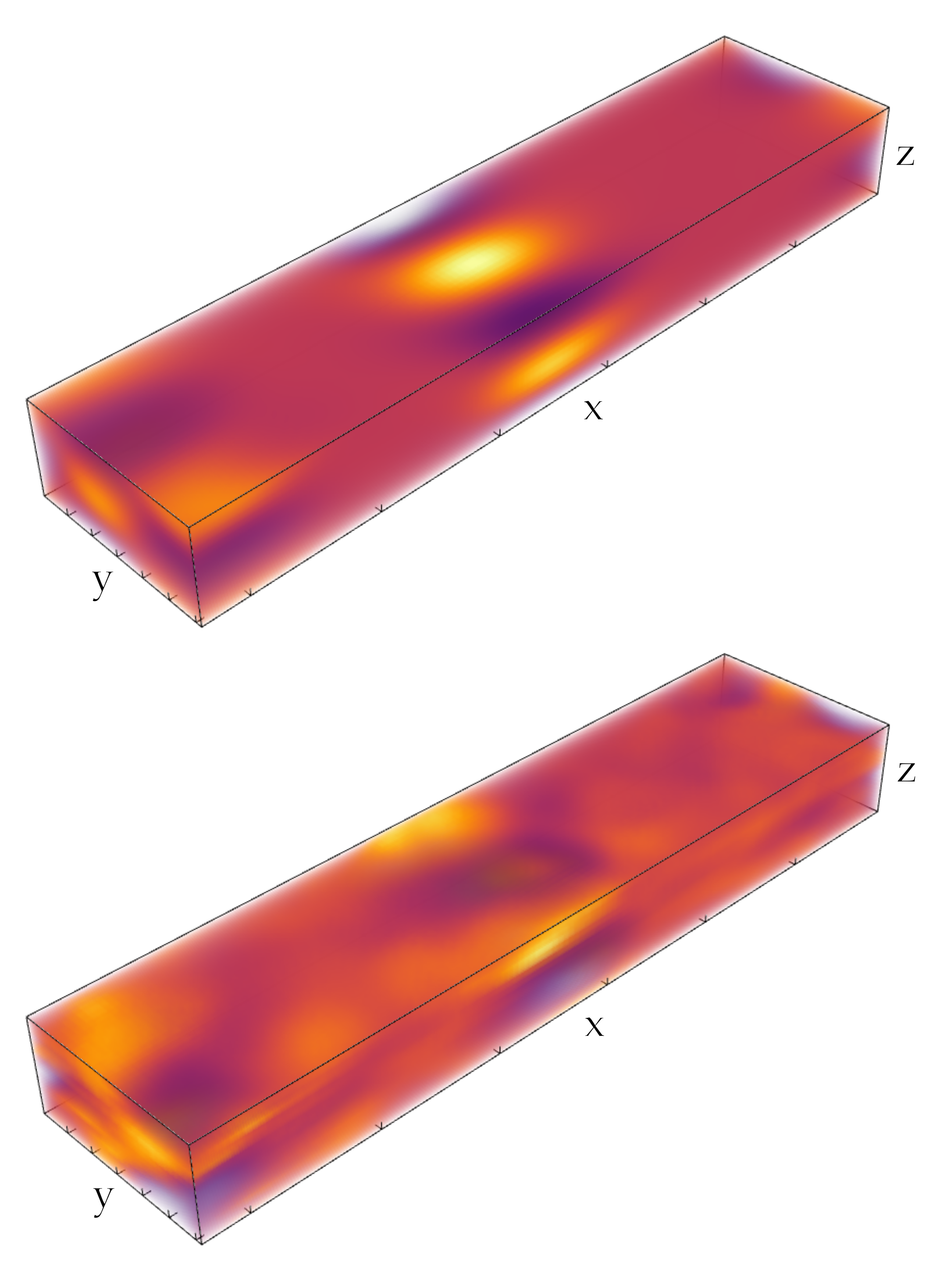}
	\caption{Snapshots of electrostatic potential in simulation of an unstable ITG modes with nonlinear treatment of
		temperature gradient at two different times during the linear (up) and nonlinear (down) phase \label{fig_snapshots_itg_real_gradient}}
\end{figure}
A snapshot of the perturbation in the linear phase is displayed in figure \ref{fig_snapshots_itg_real_gradient}. The perturbation exhibits two peaks situated at the maxima of the normalized gradient $\frac{\partial x T(x)}{T(x)}$. To determine the wavenumber, a sine wave is fitted to the envelope of the unstable within the full width half maximum (FWHM) of the perturbation (in x-direction), resulting in a wavenumber of approximately $k_x \approx 0.44$.  To facilitate a comparison between the nonlinear gradient and the analytical calculation, the mean of the gradient is computed across the mode profile.
Upon determining the wavenumber and effective gradient, the solution of the dispersion relation is obtained from the analytical expression in equation (\ref{eq_dispersion_relation}). The anticipated complex frequency for the fastest-growing mode is $	\omega = 0.01850 + 0.00420 i$.
\begin{figure}
	\centering
	\includegraphics[width=0.4\textwidth ]{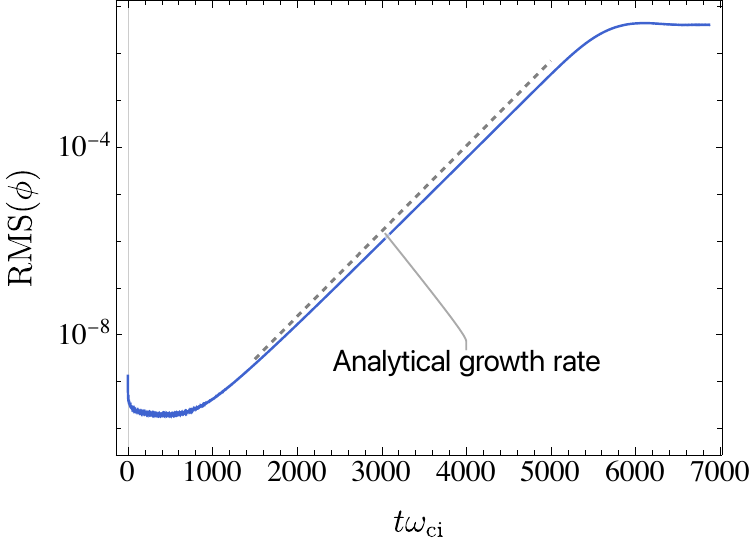}
	\caption{Root mean square of electrostatic potential compared to expected growth rate \label{fig_growth_nonlin_gradient}}
\end{figure}
Figure \ref{fig_growth_nonlin_gradient} displays the root-mean-square (RMS) of the electrostatic potential on a logarithmic scale (shown in blue), alongside the expected linear growth rate represented by a dashed gray line. The results indicate that the analytically predicted growth rate is slightly higher than the expected value. The growth rate and frequency for the simulation are derived from the Fourier-transformed electrostatic potential $\phi(\vec k ,t)$, resulting in
\begin{align}
	\omega_{\text{BSL}} = 0.01809 + 0.00409 i.
\end{align}
The growth rate and frequency exhibit an error of approximately 2\% when compared to the analytically calculated values.

After the linear phase, the simulation gradually reaches saturation around $t \approx 6000$ and settles into a fully saturated turbulent state (compare figure \ref{fig_snapshots_itg_real_gradient}(below)). This simulation marks the first of its kind in modeling developed ITG turbulence within a 6D kinetic model with a nonlinear treatment of the gradients.
\section{Summary}
Six-dimensional kinetic simulations, which enable the representation of modes with frequencies around the Larmor frequency (such as ion Bernstein waves), serve as a valuable tool for investigating physics beyond conventional gyrokinetic models. The development of our semi-Lagrangian solver for the Vlasov system marks the initial stride in constructing a comprehensive kinetic simulation code for plasma simulations at all tokamak relevant parameters.

The verification tests conducted demonstrated excellent agreement in the growth rate and frequency for the gyrokinetic modes induced by the Ion Temperature Gradient (ITG) instability, with the values computed from the analytical dispersion relation (\ref{eq_dispersion_relation}). This consistency extends across a broad range of wave numbers. Additionally, we established that the code accurately reproduces the quasi-linear fluxes throughout the domain. In the course of this, a novel formulation of the energy flux in the 6D kinetic system has been developed, representing a sum of its distinct components (namely, Poynting flux, $\vec E\times \vec B$ energy flux, and stress-induced energy flux). The explicit expression for the energy flux establishes a connection to gyrokinetic theory, as the $\vec E \times \vec B$ heat flux can be recognized as one of its components. Additionally, the individual contributions can be analytically computed, facilitating a comparison with the simulation results.
Beyond the linear verification of our model, we have conducted simulations extending well into the nonlinear phase. Comparisons with the nonlinear gyrokinetic code CGYRO indicate that the code accurately depicts the nonlinear saturation of a slab ITG instability, yielding consistent saturation amplitudes. Examination of the saturation phase and the ensuing turbulent state has led to the identification of the excitation of high-frequency modes. The gyrokinetic model, by definition, does not account for these excited ion Bernstein waves. The potential impact of these modes on the saturation process and energy transport remains unexplored. Some studies from the 1990s suggested that these waves could suppress turbulence levels in the plasma edge \cite{craddock_theory_1994,crawford_absolute_1965}. Additionally, alongside simulations employing a Boussinesq limit of the gradients, we have demonstrated the reproducibility of ITG simulations with a nonlinear treatment of the gradients. 
The possession a tool that enables the simulation of such modes presents a valuable opportunity to uncover new physics in regimes beyond commonly utilized models.

\begin{acknowledgments}
  This work has been carried out partly within the framework of the EUROfusion
Consortium, funded by the European Union via the Euratom Research and Training
Programme (Grant Agreement No 101052200 – EUROfusion). Support has also been
received by the EUROfusion High Performance Computer (Marconi-Fusion). Views
and opinions expressed are however those of the author(s) only and do not
necessarily reflect those of the European Union or the European
Commission. Neither the European Union nor the European Commission can be held
responsible for them.  Numerical simulations were performed at the
MARCONI-Fusion supercomputer at CINECA, Italy and at the HPC system at xthe Max
Planck Computing and Data Facility (MPCDF), Germany.
\end{acknowledgments}

\bibliographystyle{abbrv}

\end{document}